\documentstyle[prl,preprint,aps]{revtex}
\begin{document}
\draft
\title{Quantum chaos at the Kinetic Stage of Evolution}
\author{A. Ugulava, L. Chotorlishvili, K. Nickoladze.}
\address{Tbilisi State University,Department of Physics,
 Chavchavadze av. 3, 0128 Tbilisi, Georgia}
\date{\today}
\maketitle

\begin{abstract}

The mathematical pendulum is the simplest system having all the
basic properties inherent in dynamic stochastic systems. In the
present paper we investigate the mathematical pendulum with the
aim to reveal the properties of a quantum analogue of dynamic
stochasticity or, in other words, to obtain the basic properties
of quantum chaos.

It is shown that a periodic perturbation of the quantum pendulum
(similarly to the classical one) in the neighbourhood of the
separatrix can bring about irreversible phenomena. As a result of
recurrent passages between degenerate states, the system gets
self-chaotized and passes from the pure state to the mixed one.
Chaotization involves the states, the branch points of whose
levels participate in a slow "drift" of the system along the
Mathieu characteristics this "drift" being caused by a slowly
changing variable field. Recurrent relations are obtained for
populations of levels participating in the irreversible evolution
process. It is shown that the entropy of the system first grows
and, after reaching the equilibrium state, acquires a constant
value.
\end{abstract}
\pacs{PACS number: 87.22.Jb}

\section{Introduction. Formulation of the problem}

Dynamic stochasticity is directly connected with the assumption
that classical equations of motion may contain nonlinearities
which arise when the (exponential) repulsion of phase trajectories
occurs at a sufficiently quick rate. In the case of quantum
consideration, the dynamics of a system is described by a wave
function that obeys a linear equation, while the notion of a
trajectory is not used at all. Hence, at first sight it seems
problematic of find out the quantum properties of systems whose
classical consideration reveals their dynamic stochasticity. A
quantum analogue of classical stochastic motion is usually called
quantum chaos.

On the other hand, it is of practical interest to investigate
parametrically dependent Hamiltinians $H(Q,P,l)$, where $(Q,P)$ is
the set of canonical coordinates, $l$ is the parameter describing
how the system is related to the external field. The interest in
such systems is explained by their use in the study of quantum
points and other problems of mesoscopic physics [1].

In most of the papers that deal with parametrically dependent
systems, their authors consider the following situation. For
$l=0$, the Hamiltonian $H(Q,P,0)$ is exactly integrable. As $l$
increases, the Hamiltonian $H(Q,P,l)$ becomes nonintegrable and,
for a certan value of $l_0$, solutions of the classical equations
corresponding to $H(Q,P,l_0)$ become chaotic. In the case of
quantum consideration, eigenvalues $E_n (l_0)$ and eigenfunctions
$\psi _n (l_0)$ are found in the above-mentioned area of parameter
values by using the method of numerical diagonalization. In that
case, we show interest in the dependence of the parametrical
kernel $P(n/m)=|<\psi _n (l_o+\delta l)|\psi_m(l_o)>|^2$ on a
parameter displacement $\delta l\ll l$. The value $P(n/m)$,
averaged statistically over states $n$,
$\overline{P(n/m)}=\overline{P(n/n+r)}=P(r)$ can be interpreted as
the local density of states. The introduction of $P(r)$ means that
we pass from the quantum-mechanical description to the quantum
statistical description [2 - 5] carried out by an intuitive
reasoning.

In the problems considered in the above-listed papers, the
Hamiltonian $H(Q,P,l)$ displays chaos for both parameter values
$l=l_0$ and $l=l_0 + \delta l$.

As different from these papers, in the present paper we
investigate the situation, in which the Hamiltonian $H(Q,P,l)$ is
integrable and becomes nonintegrable after adding a strictly
periodic perturbation $\delta l(t)$. As the basic Hamiltonian we
take the Hamiltonian of the mathematical pendulum (universal
Hamiltonian).

As is known, the Schr\"{o}dinger quantum-mechanical equation for
the universal Hamiltonian is written in the form of the Mathieu
equation. The Mathieu-Schr\"{o}dinger equation for an atom, which
is under the action of optical pumping in the area of large
quantum numbers, was for first time obtained by G. Zaslavsky and
G. Berman [6]. These authors also performed analysis of
quasiclassical states of the Mathieu-Schr\"{o}dinger equation [6].

The main objective of the present paper is to investigate the
behavior of the quantum mathematical pendulum in the area of
dynamic stochasticity parameters. As is known [7], this area,
called the stochastic layer, lies in the neighborhood of the
separatrix of the classical pendulum.

We show here that with the appearance of quantum chaos the pure
state passes to the mixed one. In other words, the reversible
quantum process transforms to the irreversible process of quantum
chaos which can be described by a kinetic equation. The common
feature of classical dynamic chaos and quantum chaos is, as will
be shown below, the irreversibility of their states.

\section{A parametrically dependent Hamiltonian}

After writing the stationary Schr\"{o}dinger equation
\begin{equation}
\hat{H}\psi _n=E_n\psi _n \label{1}
\end{equation}
for the universal Hamiltonian of the atom+pumping system
$$H=-\frac{\partial ^2}{\partial \varphi^2}+V,$$ $$V=l\cos
{2\varphi},$$ we come to the equation coinciding with the Mathieu
equation [8, 9]
\begin{equation} \frac{d^2\psi _n}{d\varphi ^2}+(E_n-V(l,\varphi))\psi
_n=0,  \label{2} \end{equation} where
$E_n\rightarrow\frac{8E_n}{\hbar^2\omega^\prime}$ are the
introduced dimensionless values, $l$ is the dimensionless
amplitude of pumping, $\omega^\prime=\frac{d\omega(I)}{dI}$ is the
derivative of nonlinear oscillation frequency $\omega(I)$ with
respect to the action $I$ [9].

The Mathieu-Schr\"{o}dinger equation is characterized by a
specific dependence of the spectrum of eigenvalues $E_n(l)$ and
eigenfunctions $\psi_n(\varphi,l)$ on the parameter $l$ (see
Fig.1). On the plane $(E,l)$ with the spectral characteristics
(so-called Mathieu characteristics [9]) of the problem, this
specific feature manifests itself in the alternation of areas of
degenerate ($G_\mp$) and nondegenerate ($G$) states (see [10],
Figs.3, 4). The boundaries between these areas pass through the
branch points of energy therms $E_n(l)$.

Degenerate and nondegenerate states of the quantum mathematical
pendulum were established by studying the symmetry properties of
the Mathieu-Schr\"{o}dinger equation. In [10], by using the
symmetry properties of the Mathieu-Schr\"{o}dinger equation and
applying the group theory methods, the eigenvalues for each of the
areas $G_+$, $G_-$, $G$ were found: $$G_-\rightarrow \psi^\pm
_{2n+1}(\varphi)=\frac{\sqrt{2}}{2} (ce_{2n+1}(\varphi)\pm
ise_{2n+1}(\varphi)),~~~~~~~~(3G_-)$$ $$\psi^\pm
_{2n}(\varphi)=\frac{\sqrt{2}}{2} (ce_{2n}(\varphi)\pm
ise_{2n}(\varphi)),$$ \begin{equation} G\rightarrow
ce_{2n}(\varphi); ce_{2n+1}(\varphi); se_{2n}(\varphi);
se_{2n+1}(\varphi),~~~~~~~~~~~~~~(3G) \label{3} \end{equation}
$$G_+\rightarrow \xi^\pm _{2n}(\varphi)=\frac{1}{\sqrt{2}}
(ce_{2n}(\varphi)\pm ise_{2n+1}(\varphi)),~~~~~~~~~~~~~(3G_+)\\$$
$$\zeta^\pm _{2n+1}(\varphi)=\frac{1}{\sqrt{2}}
(ce_{2n+1}(\varphi)\pm ise_{2n+2}(\varphi)).$$ Here
$ce_n(\varphi)$ and $se_n(\varphi)$ denote the Mathieu functions
[9].

The wave functions $(3G_\pm)$ and $(3G)$ form the bases of
irreducible representations of the respective groups. Each of the
four functions $(3G)$ forms a one-dimensional irreducible
representation of the Klein group $V$, while the functions
$\psi^\pm_{2n+1}(\varphi),\psi^\pm_{2n}(\varphi)$ from $(3G_-)$
and $\xi^\pm_{2n}(\varphi), \zeta^\pm_{2n+1}(\varphi)$ from
$(3G_+)$ form the two-dimensional irreducible representations of
two invariant subgroups of the group $V$ [10].

Let us assume that the pumping amplitude is modulated by a slowly
changing electromagnetic field. The influence of modulation can be
taken into account by making a replacement in the
Mathieu-Schr\"{o}dinger equation
\begin{equation} l(t)\rightarrow l_o + \Delta l\cos \nu t, \label{4}
\end{equation}
where $\Delta l$ is the modulation amplitude expressed in
dimensionless units, $\nu$ is the modulation frequency.

We assume that a gradual change of $l(t)$ may involve some $N$
branch points on the left and on the right side of the separatrix
(Fig.1):
\begin{equation}
\Delta l \geq |l^n_+-l^n_-|,~~~n=1,2,\ldots N. \label{5}
\end{equation}
After making replacement (4) in the universal Hamiltonian, we
obtain $$\hat{H}=\hat{H}_o+\hat{H}^\prime(t),\\$$
\begin{equation}
\hat{H}_o=-\frac{\partial ^2}{\partial \varphi^2}+l_o\cos
2\varphi, \label{6}
\end{equation}
$$~~~~~~~~~~~~~~~~~~~~~~~~~~~~~~~~~~~~~~~~~~~\hat{H}^\prime
(t)=\Delta l\cos 2\varphi \cos \nu
t.~~~~~~~~~~~~~~~~~~~~~~~~~~~~~~~~~~~~~~~~~(6^\prime)\\$$

Simple calculations show that the matrix elements of perturbation
$\hat{H}^\prime (t)$ with respect to the wave functions $(3G)$ of
the nondegenerate area $G$ are equal to zero
\begin{equation} \langle
ce_n|\hat{H}^\prime(t)|se_n\rangle \sim\Delta
l\int\limits^{2\pi}_o ce_n(\varphi)\cos2\varphi
se_n(\varphi)d\varphi =0, \label{7}
\end{equation}
where $n$ is any integer number. Therefore perturbation
$(6^\prime)$ cannot bring about passages between nondegenerate
levels.

The interaction $\hat{H}^\prime (t)$, not producing passages
between levels, should be inserted in the unperturbed part of the
Hamiltonian. The Hamiltonian obtained in this manner can be
considered as slowly depending on time.

Thus, in the nondegenerate area $G$ the Hamiltonian can be written
in the form $$\hat{H}=-\frac{\partial ^2}{\partial \varphi
^2}+l(t)\cos 2\varphi,$$
\begin{equation}
l(t)=l_o + \Delta l \cos \nu t. \label{8}
\end{equation}

As has been mentioned above, to different areas on the plane
$(E,l)$ we can assign different eigenfunctions $(3G_-)$, $(3G)$,
$(3G_+)$. Because of the modulation of the parameter $l(t)$ the
system passes from one area to another, getting over the branch
points.

\section{Passage from the quantum-mechanical description
to the kinetic description. Irreversible phenomena}

As different from the nondegenerate states area $G$, in the areas
of degenerate states $G_-$ and $G_+$, the nondiagonal matrix
elements of perturbation $\hat{H}^\prime(t)$ $(6^\prime)$ are not
equal to zero. For example, if we take the matrix elements with
respect to the wave functions $\psi^\pm_{2n+1}$ (see $(3G_-)$),
then for the left degenerate area $G_-$ it can be shown that
\begin{equation}
H^\prime_{+-}=H^\prime_{-+}=<\psi^+_{2n+1}|\hat{H^\prime(t)}|\psi^-_{2n+1}>
\sim\Delta l
\int\limits_{o}^{2\pi}\psi^+_{2n+1}\psi^{-\ast}_{2n+1}\cos
2\varphi d\varphi \not= 0. \label{9}
\end{equation}
Note that the value $H^\prime_{+-}$ has order equal to the pumping
modulation depth $\Delta l.$

Analogously to (9), we can write an expression for even $2n$
states as well.

An explicit dependence of $\hat{H}^\prime(t)$ on time given by the
factor $\cos \nu t$ is assumed to be slow as compared with the
period of passages between degenerate states that are produced by
the nondiagonal matrix elements $H^\prime_{+-}$. Therefore below
the perturbation $H^\prime_{+-}$ will be assumed to be the
time-independent perturbation that can bring about passages
between degenerate states.

Thus, in a degenerate area the system may be in the time-dependent
superpositional state
\begin{equation}
\psi_{2n}(t)=C^+_n(t)\psi^+_{2n}+C^-_n(t)\psi^-_{2n}. \label{10}
\end{equation}
The probability amplitudes $C^{\pm}_n(t)$ are defined by means of
the fundamental quantum-mechanical equation expressing the
casuality principle [11]. We write such equations for a pair of
doubly degenerate states:
$$-i\hbar\frac{dC^+_n}{dt}=(E_{on}+H^\prime
_{++})C^+_n+H^\prime_{+-}C^-_n,\\$$
\begin{equation}
-i\hbar\frac{dC^-_n}{dt}=H^\prime _{+-}C^+_n+(E_{on}+H^\prime
_{--})C^-_n. \label{11}
\end{equation}
Let us solve system (11). In our case it can be assumed that
$H^\prime_{++}=H^\prime_{--}$ and $H^\prime_{+-}=H^\prime_{-+}$.
Let us investigate changes that occurred in the state of the
system during time $\Delta T$ while the system was in the area
$G_-$, assuming that $\Delta T$ is a part of the modulation period
$T,\Delta T\leq T$.

For arbitrary initial values system (11) has the solution
$$C^+_n(t)=e^{\frac{i}{\hbar}Et}(C_+\cos(\frac{H^\prime}{\hbar}t)+iC_-\sin
(\frac{H^\prime}{\hbar}t)),$$
\begin{equation}
C^-_n(t)=e^{\frac{i}{\hbar}Et}(C_-\cos(\frac{H^\prime}{\hbar}t)+iC_+\sin
(\frac{H^\prime}{\hbar}t)), \label{12}
\end{equation}
where we redenote $E_o+H^\prime_{++}\rightarrow
E,H^\prime_{+-}\rightarrow H^\prime$. A slow dependence of the
interaction $\hat{H}^\prime(t)$ (6) on time can be taken into
account in (12) if we use the replacement $H^\prime\rightarrow
H^\prime \cos \nu t$.

Let motion begin from the state $\psi^-_{2n}$ of the degenerate
area. Then as the initial conditions we take
\begin{equation}
C^-_n(0)=1,~~~C^+_n(0)=0. \label{13}
\end{equation}

Having substituted (13) into (12), for the amplitudes $C^\pm_n(t)$
we obtain $$C^+_n(t)=iexp(\frac{i}{\hbar}Et)\sin\omega t,$$
\begin{equation}
C^-_n(t)=exp(\frac{i}{\hbar}Et)\cos \omega t, \label{14}
\end{equation}
where $\omega = \frac{2\pi}{\tau}=\frac{H^\prime}{\hbar}$ is the
frequency of passages between degenerate states, $\tau$ is the
passage time.

Note that the parameter $\omega$, which is connected with the
modulation depth $\Delta l$, has (like any other parameter) a
certain small error $\delta \omega$, which during the time of one
passage $t\sim 2\pi /\omega$, leads to an insignificant correction
in the phase $2\pi(\delta\omega/\omega)$. But during the time
$t\sim\Delta T$, there occur a great number of oscillations (
phase incursion takes place) and, in the case $\Delta T\gg \tau$,
a small error $\delta \omega$ brings to the uncertainty of the
phase $\sim \Delta T \delta \omega$ which may have order $2\pi$.
Then we say that the phase is self-chaotized.

Let us introduce the densitity matrix averaged over a small
dispersion $\delta \omega$:
\begin{eqnarray} &\rho^{+-}_n(t)=\left(\begin{array}{cc}
W^+_n(t)&iF_n(t)\\ -iF^\ast_n(t)&W^-_n(t)\\
\end{array}\right),\,&
\end{eqnarray}
where $W^\pm_n(t)=\overline{|C^\pm_n(t)|^2},~~
F_n(t)=\overline{C^+_n(t)C^{-\ast}_n(t)}$. The overline denotes
the averaging over a small dispersion $\delta \omega$
\begin{equation}
\overline{A(\omega,t)}=\frac{1}{2\delta
\omega}\int\limits^{\omega+\delta \omega}_{\omega-\delta
\omega}A(x,t)dx \label{16}
\end{equation}

To solve (16) we can write that
\begin{equation}
W^+_n(t)=\overline{\sin^2\omega
t},~~W^-_n(t)=\overline{\cos^2\omega
t},~~F_n(t)=\frac{1}{2}\overline {\sin2\omega t}. \label{17}
\end{equation}

After a simple integration of the averaging (16), for the matrix
element (17) we obtain $$W^\pm_n(t)=\frac{1}{2}(1\mp f(2\delta
\omega t)\cos 2\omega t),$$
\begin{equation}
F_n(t)=F^\ast_n(t)=\frac{1}{2}f(2\delta \omega t)\sin 2\omega t,
\label{18}
\end{equation}
$$f(2\delta \omega t)=\frac{\sin2\delta \omega t}{2\delta \omega
t}.$$

At small values of time
$t\ll\overline{\tau}~(\overline{\tau}=2\pi/\delta \omega)$,
insufficient for self-chaotization  $(f(2\delta \omega t)\approx
1)$, we obtain $$W^+_n(t \ll\overline{\tau})=\sin^2 \omega
t,~~W^-_n(t \ll\overline{\tau})=\cos^2\omega
t,~~F_n(t\ll\overline{\tau})=\frac{1}{2}\sin2\omega t.$$

Comparing these values with the initial values (17) of the density
matrix elements, we see that the averaging procedure (16), as
expected, does not affect them. Thus, for small times we have
\begin{eqnarray}
&\rho^{+-}_n(t\ll\overline{\tau})=\left(\begin{array}{cc}
\sin^2\omega t&\frac{i}{2}\sin 2\omega t\\ \frac{-i}{2}\sin2\omega
t&\cos^2\omega t\\
\end{array}\right).\,&
\end{eqnarray}

One can easily verify that matrix (19) satisfies the condition
$\rho^2_n(t\ll\overline{\tau})=\rho_n(t\ll\overline{\tau})$, which
is a necessary and sufficient condition for the density matrix of
the pure state.

For times even smaller than $t\ll \tau\ll \overline{\tau}$, when
passages between degenerate states practically fail to occur, by
taking the limit $\omega t\ll 1$ in (19), we obtain the following
relation for the density matrix:
\begin{equation} \rho^{+-}_n(t=0)=\rho^{+-}_n(t\ll\tau)=\left(\begin{array}{cc}
0&0\\ 0&1
\end{array}\right).\,
\end{equation}
This relation corresponds to the initial relation (13) when the
system is in the eigenstate $\psi^-_{2n}$. Let us now investigate
the behavior of the system at times $t\geq\overline{\tau}$ when
the system gets self-chaotized.

On relatively large time intervals $t\geq \overline{\tau}$, in
which the self-chaotization of phases  takes place, for the matrix
elements we should use general expressions (18). The substitution
of these expressions for the matrix elements (18) into the density
matrix (15) gives
\begin{eqnarray} &\rho^{+-}_n(t)=\frac{1}{2}\left(\begin{array}{cc}
1-f(2\delta \omega t)\cos 2\omega t&f(2\delta \omega t)\sin2\omega
t\\ -if(2\delta \omega t)\sin2\omega t&1+f(2\delta \omega
t)\cos2\omega t\\
\end{array}\right).\,&
\end{eqnarray}
Hence, for times $t\geq\overline{\tau}$ during which the phases
get completely chaotized, after passing to the limit $\delta
\omega t\gg1$ in (21), we obtain
\begin{eqnarray}
&\rho^{+-}_n(t\gg\overline{\tau})=\frac{1}{2}\left(\begin{array}{cc}
1-O(\epsilon)&iO(\epsilon)\\-iO(\epsilon)&1+O(\epsilon)\\
\end{array}\right),\,& \end{eqnarray}
where $O(\epsilon)$ is an infinitesimal value of order
$\epsilon=\frac{1}{2\delta \omega t}$.

The state described by the density matrix (22) is a mixture of two
quantum states $\psi^+_{2n}$ and $\psi^-_{2n}$ with equal weights.
The comparison of the corresponding matrix elements of matrices
(22) and (21) shows that they differ in the terms that play the
role of quickly changing fluctuations. When the limit is
$t\gg\overline{\tau}$, fluctuations decrease as
$\sim\frac{1}{2\delta \omega t}$ (see Fig.2 and 3).

Thus the system, which at the time moment $t=0$ was in the pure
state with the wave function $\psi^-_{2n}$ (20), gets
self-chaotized with a lapse of time $t\gg\overline{\tau}$ and
passes to the mixed state (22). In other words, at the initial
moment the system had a certain definite "order" expressed in the
form of the density matrix $\rho^{+-}(0)$ (20). With a lapse of
time the system got self-chaotized and the fluctuation terms
appeared in the density matrix (21). For large times
$t\gg\overline{\tau}$ a new "order" looking like a macroscopic
order is formed, which is defined by matrix (22).

After a halfperiod the system passes to the area  of nondegenerate
states $G$ (20). In passing through the branch point, there arise
nonzero probabilities for passages both to the state $ce_{2n}$ and
to the state $se_{2n}$. Both states $\psi^+_{2n}$ and
$\psi^-_{2n}$ will contribute to the probability that the system
will pass to either of the states $ce_{2n}$ and $se_{2n}$. For the
total probability of passage to the states $ce_{2n}$ and $se_{2n}$
we obtain respectively $$P(\rho^{+-}_{2n}(t\gg \tau)\rightarrow
ce_{2n})=\frac{1}{2}\bigl|\frac{1}{\pi}\int\limits_{o}^{2\pi}\psi^+_{2n}
(\varphi)ce_{2n}(\varphi)d\varphi\bigr|^2+$$
$$+\frac{1}{2}\bigl|\frac{1}{\pi}\int\limits_{o}^{2\pi}\psi^-_{2n}(\varphi)
ce_{2n}(\varphi)d\varphi\bigr|^2=\frac{1}{2}\cdot \frac{1}{2}+
\frac{1}{2}\cdot \frac{1}{2}=\frac{1}{2},$$
$$P(\rho^{+-}_{2n}(t\gg \tau)\rightarrow
se_{2n})=\frac{1}{2}\bigl|\frac{1}{\pi}\int\limits_{o}^{2\pi}\psi^+_{2n}
(\varphi)se_{2n}(\varphi)d\varphi\bigr|^2+$$
\begin{equation}
+\frac{1}{2}\bigl|\frac{1}{\pi}\int\limits_{o}^{2\pi}\psi^-_{2n}(\varphi)
se_{2n}(\varphi)d\varphi\bigr|^2=\frac{1}{2}\cdot \frac{1}{2}+
\frac{1}{2}\cdot \frac{1}{2}=\frac{1}{2}. \label{23}
\end{equation}
Thus, in the nondegenerate area the mixed state is formed, which
is defined by the density matrix
\begin{eqnarray}
&\rho^{ik}_{2n}(t\sim\frac{T}{2}\gg\tau)=\frac{1}{2}\left(\begin{array}{cc}
1&0\\ 0&1\\
\end{array}\right),\,&
\end{eqnarray}
where $i$ and $k$ number two levels that correspond to the states
$ce_{2n}$ and $se_{2n}$.

As follows from (24), at this evolution stage of the system, the
populations of two nondegenerate levels get equalized. It should
be noted that though the direct passage (7) between the
nondegenerate levels is not prohibited, perturbation (6')
essentially influences "indirect" passages. Under "indirect"
passages we undestand a sequence of events consisting a passage
$G\rightarrow G_-$ through the branch point, a set of passages
between degenerate states in the area $G_-$, and the reverse
passage through the branch point $G_-\rightarrow G$. The
"indirect" passages ocurring during the modulation halfperiod
$T/2$ result in the equalization (saturation) of two nondegenerate
levels.

As to the nondegenerate area, the role of perturbation
$\hat{H}^\prime(t)$ in it reduces to the displacement of the
system from the left branch point to the right one.

It is easy to verify that after states (24) pass to the states of
the degenerate area  $G_+$, we obtain the mixed state which
involves four states $\xi^\pm_{2n}(\varphi)$ and
$\zeta^\pm_{2n+1}(\varphi)$ (see Fig.1).

Let us now calculate the probability of four passages from the
mixed state $\rho^{ik}_{2n}$ (22) to the states
$\xi^\pm_{2n}(\varphi)$ and $\zeta^\pm_{2n-1}(\varphi)$:
$$P(\rho^{ik}_{2n}\rightarrow
\xi^\pm_{2n})=\frac{1}{2}\bigl|\frac{1}{\pi}
\int\limits_{o}^{2\pi}(ce_{2n}(\varphi)+se_{2n}(\varphi))\xi^\pm_{2n}(\varphi)
d\varphi\bigr|^2=\frac{1}{4},$$
\begin{equation}
P(\rho^{ik}_{2n}\rightarrow
\zeta^\pm_{2n-1})=\frac{1}{2}\bigl|\frac{1}{\pi}
\int\limits_{o}^{2\pi}(ce_{2n}(\varphi)+se_{2n}(\varphi))\zeta^\pm_{2n-1}(\varphi)
d\varphi\bigr|^2=\frac{1}{4}. \label{25}
\end{equation}

As a result of these passages, in the area $G_+$ we obtain the
mixed state described by the four-dimensional density matrix
\begin{eqnarray}
&\rho^{+-}_{2n,2n+1}(t\sim
T\gg\tau)=\frac{1}{4}\left(\begin{array}{cccc} 1&0&0&0\\ 0&1&0&0\\
0&0&1&0\\ 0&0&0&1\\
\end{array}\right),\,&
\end{eqnarray}
where the indices of the density matrix (26) show that the
respective matrix elements are taken with respect to the wave
functions $\xi^\pm_{2n}(\varphi)$ and $\zeta^\pm_{2n+1}(\varphi)$
of degenerate states of the area  $G_+$.

It is easy to foresee a further evolution course of the system. At
each passage through the branch point, the probability that an
energy level will get populated is equally divided between
branched states. We can see the following regularity of the
evolution of populations for the next time periods.

After odd halfperiods, the population of any $n-$th nondegenerate
level is defined as an arithmetic mean of its population and the
population of the nearest upper level, while after even
halfperiods -- as an arithmetic mean of its population and the
nearest lower level. This population evolution rule can be
represented both in the form of Table 1 and in the form of
recurrent relations
$$P[n,2k]=P[n+1,2k]=\frac{1}{2}(P[n,2k-1]+P[n+1,2k-1]),\\$$
\begin{eqnarray}
P[n+1,2k+1]=P[n+2,2k+1]=\frac{1}{2}(P[n+1,2k]+P[n+2,2k]),
\label{27}
\end{eqnarray}
where $P[n,k]$ is the population value of the $n-$th level after
time $k\frac{T}{2}$, where $k$ is an integer number. The creeping
of populations among nondegenerate levels is illustrated in Fig.4.

The results of numerical calculations by means of formulas (27)
are given in Fig.5 and Fig.6. Fig.5 shows the distribution of
populations of levels $P(n)$ after a long time $t\gg T$ when the
population creeping occurs among levels, the number of which is
not restricted by (5). Let us assume that at the initial time
moment $t=0$, only one $n_0-$th level is populated with
probability $P(n_0)=1$. According to the recurrent relations (27),
with a lapse of each period $T$ "indirect" passages will result in
the redistribution of populations among the neighboring levels so
that, after a lapse of time $t=k\tau\gg T$, populations of the
extreme levels will decrease according to the law [10] $$P(n_o\pm
k)\sim\frac{1}{2^k}.$$

If the number $N$ of levels defined by condition (5) is finite,
then, after a lapse of a long time, passages will result in a
stationary state in which all $N$ levels are populated with the
same probability equal to $1/N$ (see Fig.6). The distribution
obtained by us is analogous to the distribution obtained in [12,
13] in investigating the problem on a linear oscillator under the
action of an electromagnetic field in the conditions of weak
chaos.

Let us summarize the results we have obtained above using the
notions of statistical physics. After a lapse of time $\Delta T$,
that can be called the time of initial chaotization, the
investigated closed system (quantum pendulum+variable field) can
be considered as a statistical system.  At that, the closed system
consists of two subsystems: the classical variable field (6') that
plays the role of a thermostat with an infinitely high temperature
and the quantum mathematical pendulum (6). A weak (indirect)
interaction of the subsystems produces passages between
nondegenerate levels. After a lapse of time $t\gg T$ this
interaction ends in a statististical equilibrium between the
subsystems. As a result, the quantum pendulum subsystem acquires
the thermostat temperature, which in turn leads to the
equalization of level populations. The equalization of populations
usually called the saturation of passages can be interpreted as
the acquisition of an infinite temperature by the quantum pendulum
subsystem.

\section{Entropy growth of the quantum pendulum subsystem.
Variable field energy absorption}

As is known, variation or constancy of entropy can be considered
as a criterion of irreversibility and reversibility of processes
occurring in a closed system. In the case of irreversible
processes, during which the system tends to the equilibrium state,
the entropy increases, while in the equilibrium state it remains
constant.

Let us use this criterion to clarify the question of reversibility
for our problem. As is known, the entropy of an arbitrary quantum
system is defined by the operator of the density matrix
$\hat{\rho}$ [14]
\begin{equation}
S(t)=-<\overline{\hat{\rho}(t)}ln\overline{\hat{\rho}(t)}>,
\label{28}
\end{equation}
where the brackets $<\ldots>$ denote the quantum-mechanical
averaging, while the overline denote the averaging over a small
dispersion $\delta \omega$ of the passage frequency (9). In the
matrix form the right-hand side of formula (24) is written as
\begin{equation}
S(t)=-Tr(\overline{\hat{\rho}(t)}ln\overline{\hat{\rho}(t)})=-
\sum\limits_{i,k}\overline{\rho^{ik}(t)}ln\overline{\rho^{ki}(t)}.
\label{29}
\end{equation}

If initially the system is in one of degenerate states, for
example, in the pure state $\psi^-_{2n}$, then
$C^-_n(0)=1,~~C^+_n(0)=0,$
$$\rho^{+-}_n(t=0)=\left(\begin{array}{cc} 0&0\\ 0&1\\
\end{array}\right),$$
and therefore, by (29), the entropy is $S_n(t=0)=0$.

With a lapse of time $t\gg \tau$, as passages between
nondegenerate states are completed and the self-chaotization
condition $\Delta T\delta \omega\geq2\pi,$ is fulfilled, the
density matrix takes form (22). Then the substitution of the
density matrix (22) into the entropy formula (29) gives
\begin{equation}
S_n(t\sim \frac{T}{2}\gg\tau)=ln2. \label{30}
\end{equation}
Thus, on a time interval from $t=0$ to $t\leq \Delta T$, the
entropy grows $$ S_n(t\gg \tau) > S_n(t=0).$$ This proves that on
this time interval the process is irreversible.

Using analytical methods, we have succeeded in establishing only
the asymptotic value of entropy. To investigate a complete picture
of entropy change on a time interval $0\leq t\leq \Delta T$, we
use expression (15) for $\rho^{+-}_n(t)$. After substituting it
into the entropy formula we obtain
\begin{equation}
S_n(t)=-W^-_n(t) ln |W^-_n(t)| - W^+_n(t) ln |W^+_n(t)|-\pi
F_n(t). \label{31}
\end{equation}
Fig.7 shows the entropy as a function of time constructed with the
aid of (18), (31) by numerical methods.

To calculate the entropy value with the lapse of one period $T$,
we substitute matrix (26) into the entropy formula (29) and thus
obtain $$S_n(t\sim T)=ln4.$$

The state, in which all accessible levels of the subsystem are
populated with the same probability $1/N$ (Fig.6), is the
equilibrium state. The corresponding density matrix of dimension
$N$ is written as
\begin{eqnarray}
&\rho^{ik}_{2n}(t\gg T)=\frac{1}{N}\left(\begin{array}{cccccc}
1&0&.&.&.&0\\ 0&1&.&.&.&0\\ 0&0&1&.&.&0\\ .&.&.&.&.&.\\
.&.&.&.&.&.\\ 0&0&0&0&0&1\\
\end{array}\right).\,&
\end{eqnarray}
After substituting $\rho^{ik}_{2n}$ from (32) into(29), we obtain
the maximal entropy value on time intervals $t\gg T$
\begin{equation}
S_n(\infty)=S_n(t\gg T)=lnN. \label{33}
\end{equation}
Thus we see that the entropy constantly grows up to value (33) and
after that it stops to grow.

Let us now calculate the energy mean of the quantum pendulum
subsystem. It is obvious that for the average energy of the
subsystem we can write
\begin{equation}
E(t)=E_o+\sum\limits_{n=1}^N P_n(t)E_n, \label{34}
\end{equation}
where $E_o$ is the initial energy value defined by the initial
$(t=0)$ population of the levels, $P_n(t)$ is the probability that
the $n-$th level will be populated at the time moment $t$, $E_n$
is the energy value in the $n-$th state defined from the area of
nondegenerate states, In Fig. 8 we see that the subsystem energy
first grows and then becomes constant. This result can be
explained if we take into account the time dependent trend of
population changes which is defined by the recurrent relations
(27) or Table 1. At the beginning the subsystem absorbs the field
energy $(6^\prime)$ and, in doing so, performs "indirect" passages
between energy levels mostly in the upward direction. Upon
reaching the equilibrium state, in which the subsystem is
characterized by the equalization of level populations, it stops
to absorb energy.

\section{Conclusions. Analogy between the classical and the quantum
consideration}

The classical mathematical pendulum may have two oscillation modes
(rotational and oscillatory) which on the phase plane are
separated by the separatrix (see Fig.9,a).

On the plane $(E,l)$ the quantum mathematical pendulum has two
areas of degenerate states -- $G_-$ and $G_+$. Quantum states from
the area $G_-$ possesses translational symmetry in the pendulum
phase space. These states are analogous to the classical
rotational mode. Quantum states from the degenerate area  $G_+$
possess symmetry with respect to the equilibrium state of the
pendulum and therefore are analogous to the classical oscillatory
state [10]. On the plane $(E,l)$, the area  of nondegenerate
states $G$, which lies between the areas $G_-$ and $G_+$, contains
the line $E=l$ corresponding to the classical separatrix (see
Fig.9,b). If the classical pendulum is subjected to harmonically
changing force that perturbs a trajectory near to the separatrix,
then the perturbed trajectory acquires such a degree of complexity
that it can be assumed to be a random one. Therefore we say  that
a stochastic motion layer (so-called stochastic layer) is formed
in the neighborhood of the separatrix [7] (see Fig.10,a).

In the case of quantum consideration, the periodic perturbation
(6) brings about passages between degenerate states. As a result
of repeated passages, before passing to the area $G$ the system
gets self-chaotized, passes from the pure state to the mixed one
and further evolves irreversibly. While it repeatedly passes
through the branch points, the redistribution of populations by
the energy spectrum takes place. Only the levels whose branch
points satisfy condition (5), participate in the redistribution of
populations (see Fig.10,b).

\newpage

Subscripts to Figures

Fig.1 A fragment of the parameter-dependent energy spectrum of the
quantum mathematical pendulum (1).

Fig.2  Time-dependence of  the diagonal matrix element $W^+_n(t)$
of the density matrix  (15), constructed by means of formulas
(15), (18) for the parameter values $\omega=1/\tau =1, C^+_n(0)=1,
C^-_n(0)=0$. As clearly seen from the figure, the higher the
dispersion value of the parameter $\delta \omega$, the sooner the
stationary value $W^+_n(t>\overline{\tau}\sim \frac{1}{\delta
\omega})=\frac{1}{2}$ is achieved.

Fig.3  The vanishing of nondiagonal matrix elements of the density
matrix (15) with a lapse of time $t>\overline{\tau}$ while the
system remained in the degenerate area $G_-$. The graph is
constructed for the parameter values $\omega=1/\tau =1,
C^+_n(0)=1, C^-_n(0)=0$  with the aid of formulas  (15) and (18).

Fig.4  A fragment of the energy spectrum depending on the slowly
changing parameter (4) of the quantum mathematical pendulum (6).
With a lapse of time $t\gg T$ the stationary state is achieved,
for which all levels are populated with an equal probability.

Fig.5  Results of numerical calculations performed by means of
recurrent relations (27).  Formation of statistical distribution
of populations of levels $P(n)$  with a lapse of  a large
evolution time $t\approx 1000 T$ of  the system. The result shown
in this figure corresponds to the case for which the level
population creeping is not restricted by condition (5).

Fig.6  Results of numerical calculations performed by means of
recurrent relations (27). With a lapse of a large time interval
$t\sim 1000 T$ the formation of  stationary distribution of
populations among levels takes place. By computer calculations it
was found that in the stationary state all $N$ levels satisfying
condition (5) were populated with equal probability $1/N$.

Fig.7  The entropy growth graph constructed with the aid of
expression (31), using numerical methods for the parameter values
$\omega =1/\tau =1$.

Fig.8  Time-dependence of a mean energy value of the quantum
mathematical pendulum subsystem, constructed by numerical methods
using formula (34). As clearly seen from the figure, the
absorption of optical pumping energy takes place prior to reaching
the state of  statistical equilibrium.

Fig.9  Analogy between the classical and quantum considerations.
Unperturbed motion.

a) Classical case. Phase plane. Separatrix.

b) Quantum case. Specific dependence of the energy spectrum on the
parameter (Mathieu characteristics). Degenarate $G_\pm$   and
nondegenerate $G$ areas of the spectrum.

Fig.10 Analogy between the classical and quantum considerations.
Perturbed motion.

a) Classical case. Stochastic trajectories in the neighborhood of
the separatrix form the stochastic layer (cross-hatched area).

 b)Quantum case. The mixed state was formed as a result of population
of nondegenerate levels situated on both sides of the classical
separatrix.
\newpage

Subscripts to Tables

Table 1. Evolution of populations of nondegenerate levels.

This table is a logical extrapolation of the analytical results
obtained in Subsection 3. It shows how the population concentrated
initially on one level $n_o$ gradually spreads to other levels. It
is assumed that the extreme upper level $n_o+4$  and the extreme
lower level $n_o-5$ are forbidden by condition (5) and do not
participate in the process.


\newpage
\begin{references}

\bibitem{1} Y.Imry, Introduction in Mesoscopic Physics
(Oxford University Press, New York, 1997).

\bibitem{2} M.Feingold, A.Peres, Phys.Rev. A34, 591 (1986).

\bibitem{3} D.Cohen, T.Kottos, arXiv:cond-mat/0302319, v.3, 18 Mar
(2004).

\bibitem{4} D.Cohen, T.Kottos, Phys.Rev.Lett. 85, 4839 (2000).

\bibitem{5} D.Cohen, T.Kottos, Phys.Rev.E63, 36203 (2001).

\bibitem{6} G.P.Berman and G.M.Zaslavsky, Phys.Lett. 61A, 295 (1977).

\bibitem{7} R.Z.Sagdeev, D.A.Usikov, and G.M.Zaslavsky, Nonlinear
Physics (Hardwood, New-York, Acad.Publ., 1988).

\bibitem{8} A.Ugulava, L.Chotorlishvili, and K.Nickoladze, Phys.Rev.
E68, 026216 (2003).

\bibitem{9} H.Bateman and A.Erdelyi, Higher Transcendental Functions, vol.3 (MC
Grow-Hill Book Company, New York-Toronto-London, 1955).  Handbook
of Mathematical Functions, edited by M.Abramovitz, National Bureau
of Standards Applied Mathematics Series, vol.55 (Dover, New York,
1964).

\bibitem{10} A.Ugulava, L.Chotorlishvili, and K.Nickoladze, Phys.Rev. E70 p.026219
(2004).

\bibitem{11} L.D.Landau, E.M.Lifshitz, Quantum Mechanics, Nonrelativistic
Theory (Pergamon Press, Oxford, New York, 1977).

\bibitem{12} V.Ya.Demikhovski, D.I.Kamenev, Phys.Lett. A228, 391
(1997).

\bibitem{13} G.P.Berman, V.Ya.Demikhovski, D.I.Kamenev, Chaos
10, 670 (2000).

\bibitem{14} S.Fujita, Introduction to Non-equilibrium Quantum
Statistical Mechanics (W.P.Saunders Company, Philadelphia -London,
1966).
\end{references}
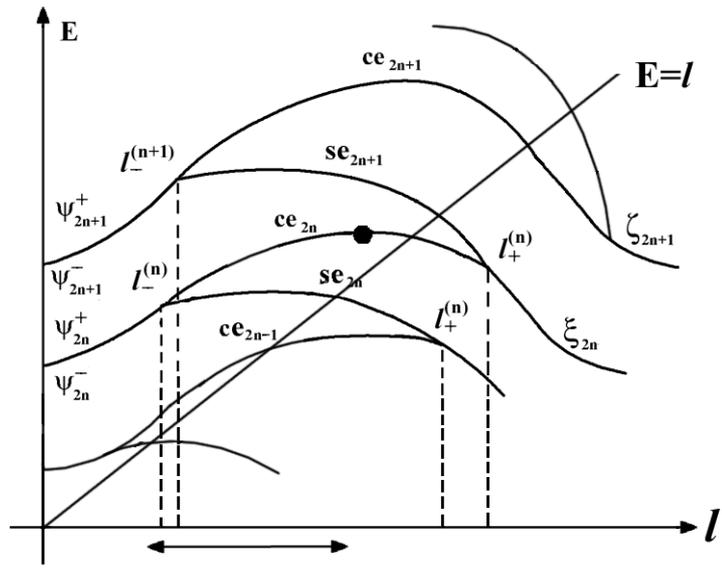
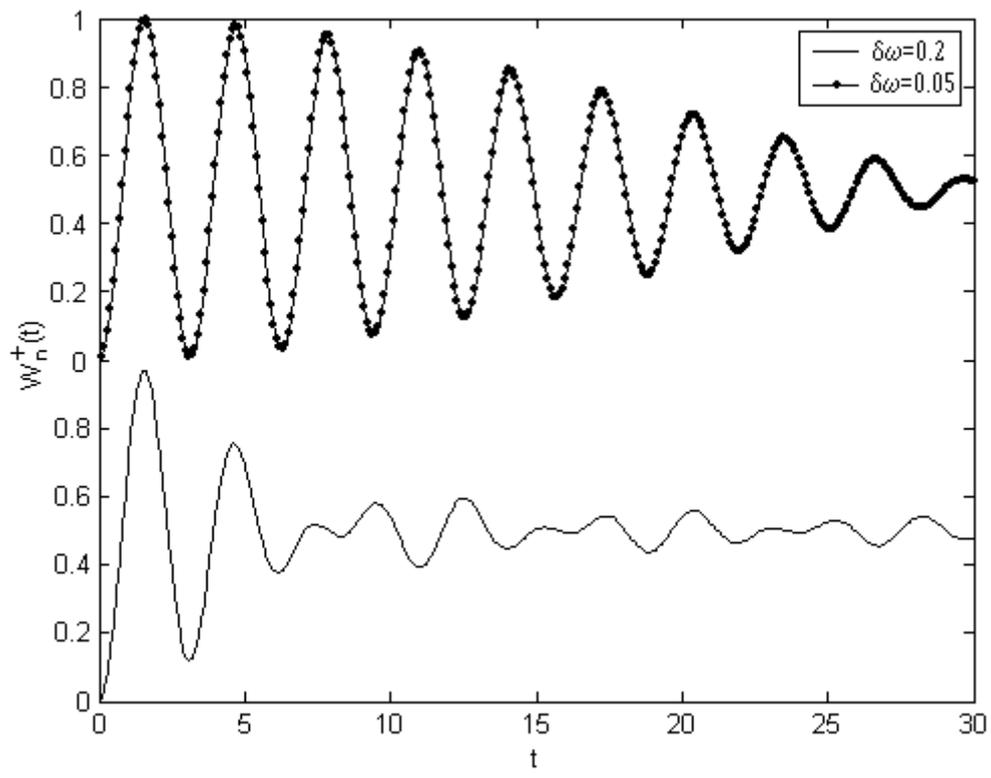
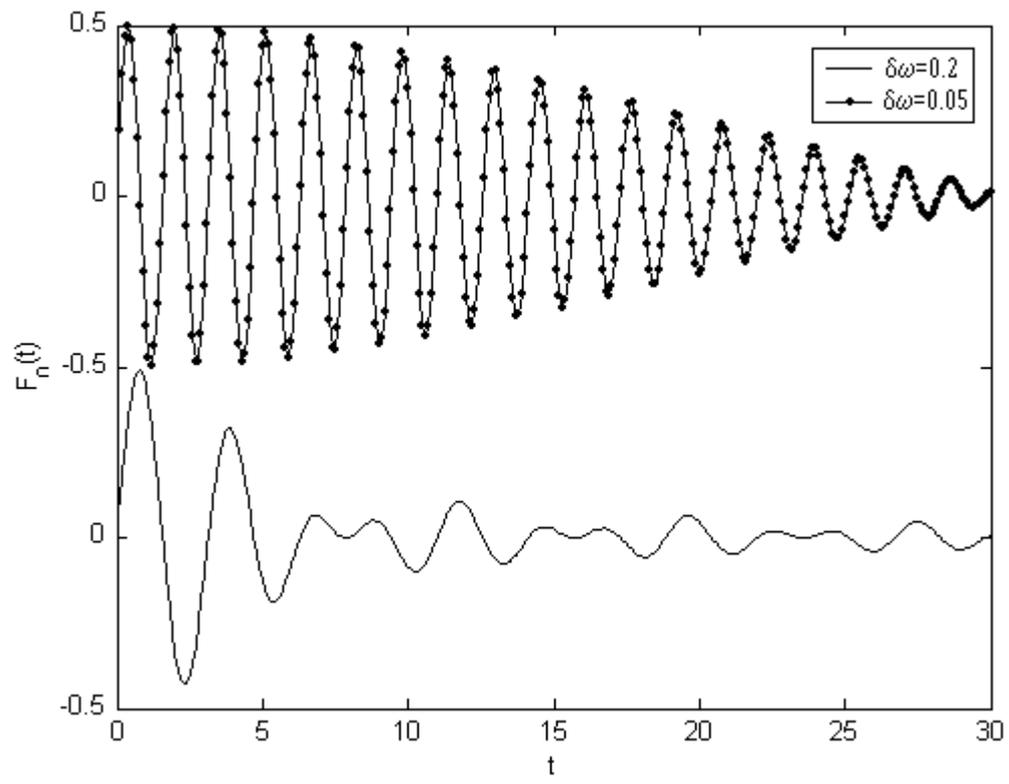
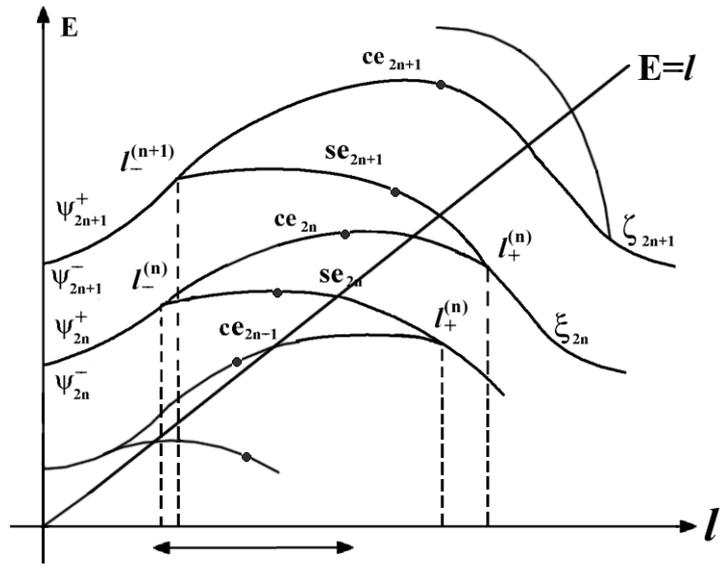
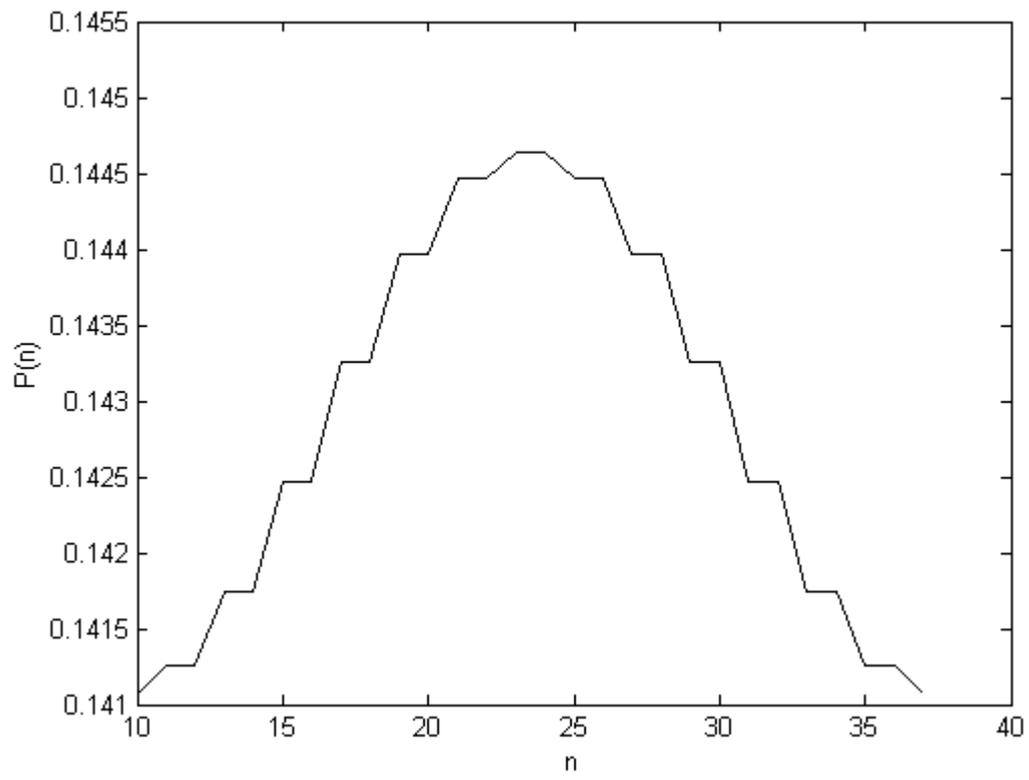
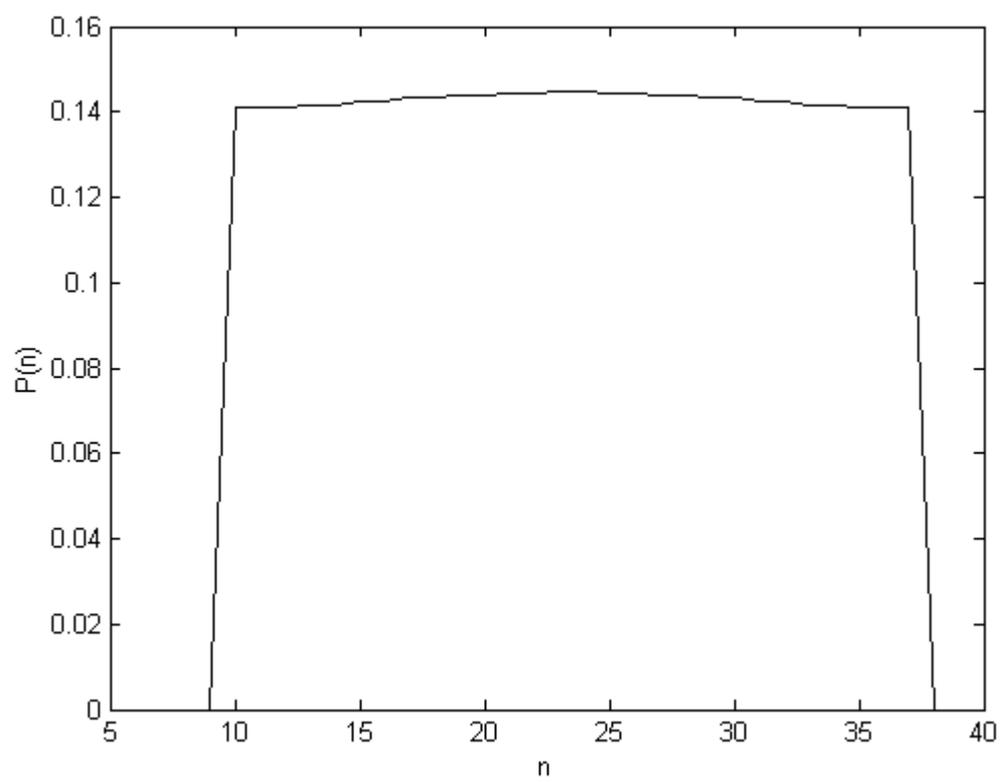
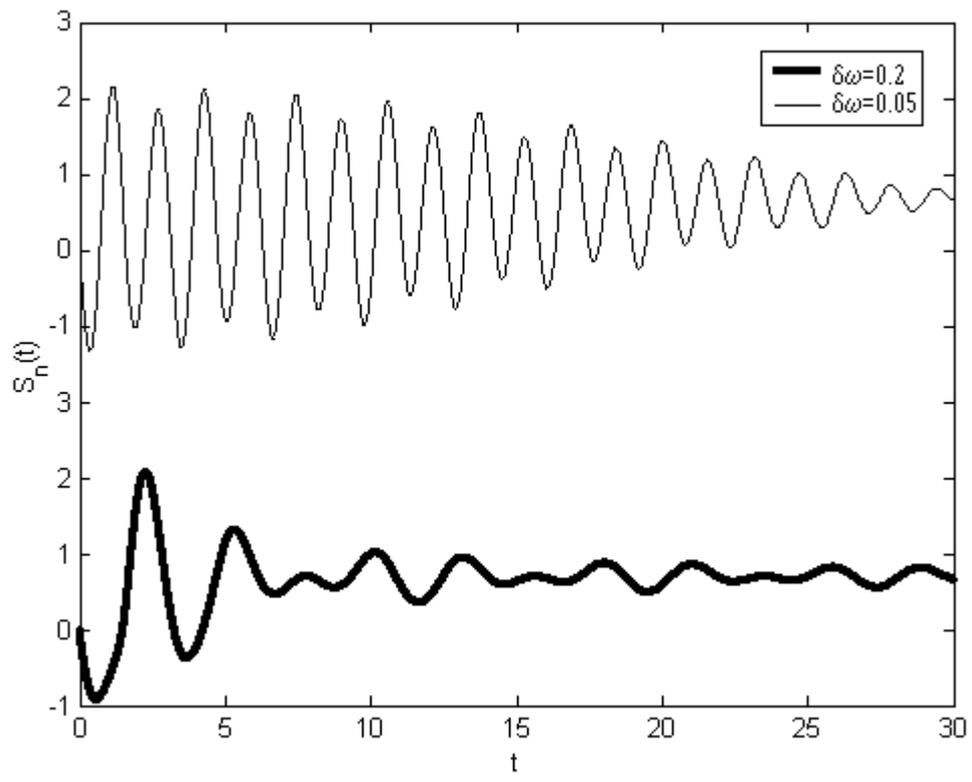
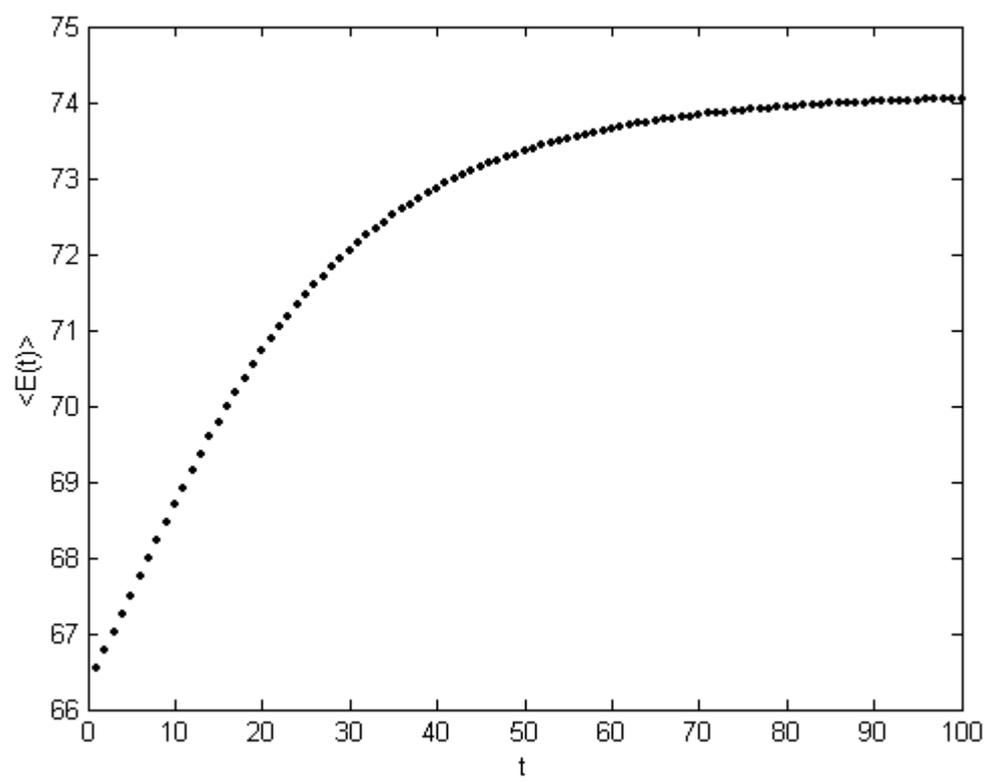
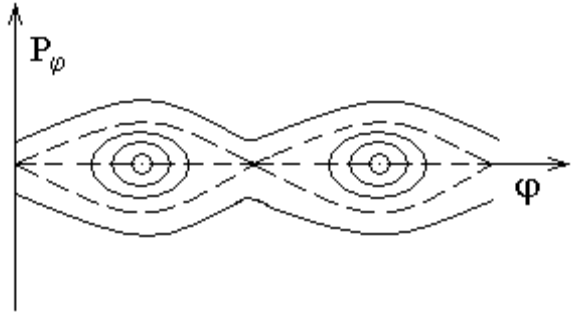
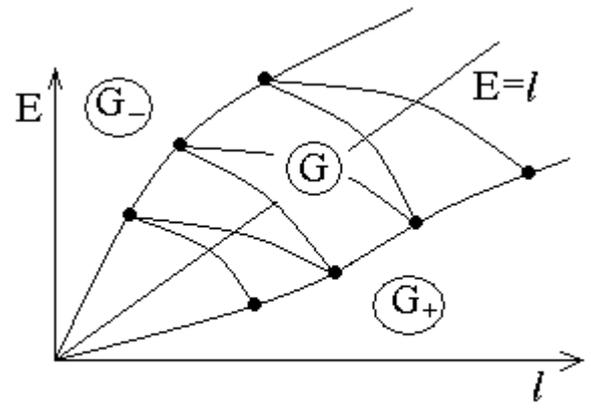
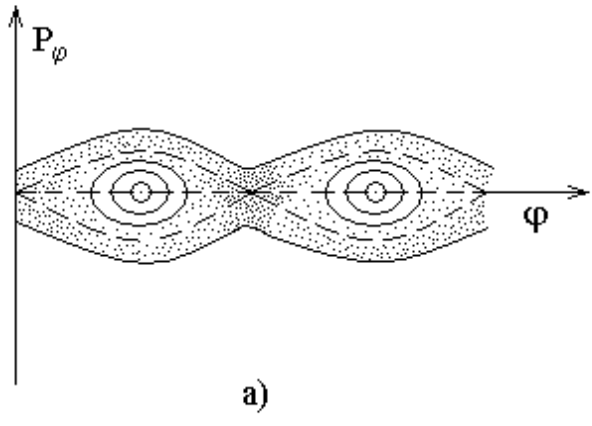
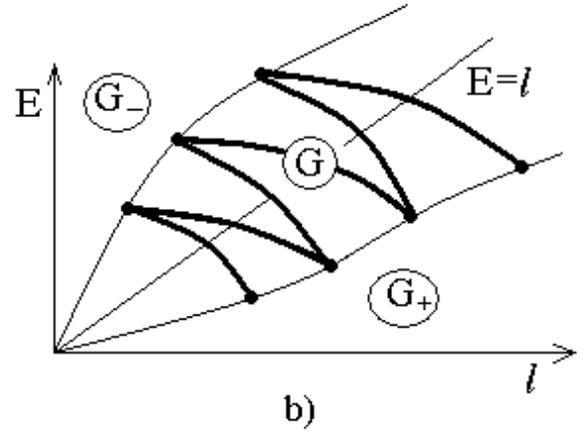
\end{document}